\documentclass[pra, aps, twocolumn, amsmath,amssymb, floatfix,nofootinbib, tightenlines,nobibnotes,longbibliography]{revtex4-2}
\usepackage{graphicx}% Include figure files
\usepackage{dcolumn}% Align table columns on decimal point
\usepackage{bm}% bold math
\usepackage{multirow}
\usepackage{graphicx}% Include figure files
\usepackage{bm}% bold math
\usepackage{graphicx, dcolumn, bm, amssymb, epstopdf, soul, amsmath}
\usepackage[normalem]{ulem}
\usepackage[caption=false]{subfig}
\usepackage{bm}
\usepackage{ragged2e}% Needed for justification=Justified (uppercase J!)
\usepackage{hyperref}

\usepackage{color}

\usepackage{ulem}
\usepackage[T1]{fontenc}
\usepackage{array}
\usepackage{booktabs}
\usepackage{multirow}
\usepackage{adjustbox}
\usepackage{etoolbox}
\AtBeginEnvironment{tabular}{\scriptsize}
\usepackage{tikz}
\usepackage{mathtools}
\newcommand\Mycomb[2][^l]{\prescript{#1\mkern-0.5mu}{}C_{#2}}

\AtBeginEnvironment{tabular}{\footnotesize}

\definecolor{auburn}{rgb}{0.63, 0.21, 0.1}

\newcommand\redsout{\bgroup\markoverwith{\textcolor{red}{\rule[0.5ex]{2pt}{0.4pt}}}\ULon}

\usepackage{amsmath}
\usepackage{mathrsfs}

\begin{document}

%Title of paper
\title{General concurrence percolation on quantum networks}

\author{Deep Nath}
\email{deepnath441@gmail.com}
\author{Soumen Roy}
\email{soumen@jcbose.ac.in}

\affiliation{Department of Physical Sciences, Bose Institute, Kolkata 700091, India}

%\email{soumen@jcbose.ac.in}

%\affiliation{Department of Physical Sciences, Bose Institute, Kolkata 700 091, India}

%\date{\today}
%\date{}

\begin{abstract}
A quantum network is a network of entangled states, which can be used to transmit quantum information. Nonmaximally entangled states are not really effective in establishing quantum communication across vast distances. Creating and maintaining a maximally entangled state over which quantum information can be transferred more effectively is difficult in experiments, as nonmaximally entangled states are typically produced there. Therefore, we usually construct a network of nonmaximally entangled states and use the underlying network structure to establish maximal entanglement between distant nodes. Various protocols leveraging interesting aspects of quantum and statistical physics are designed to achieve long-range maximally entangled states. Here, we consider two-dimensional lattices as quantum networks where edges represent nonmaximally entangled pure states. We introduce a general protocol based on bond percolation --- general concurrence percolation (GCP) --- to create a network of maximally entangled states. The associated network obtained through GCP ought to enable the achievement of long-range quantum communication. Our observations indicate that, similar to other existing protocols, GCP falls within the percolation universality class, as determined through finite-size scaling analysis performed on two-dimensional lattices. We find that the associated percolation threshold value is smaller than that of the existing protocols. We introduce and analytically solve a minimal model that renders insight into the origin of the fundamentally lower percolation threshold achieved by GCP.
\end{abstract}


\maketitle

\section{Introduction}
\label{sec:intro}

Quantum communication displays more promise than classical communication in various facets of modern science, technology, and security \cite{Buhrman1998}. Quantum communication enables us to communicate quantum information between two or more distant observers \cite{Gisin2007}. In contrast to classical information, quantum information incorporates the principles of quantum physics to provide higher storage capacity, better security, and faster transmission speed \cite{Flamini2018}. The realization of these remarkable advantages inherently relies on quantum entanglement, which serves as one of the cornerstones of quantum information \cite{Horodecki2009}. Specifically, entanglement is a fundamental resource in establishing long-range quantum communication and computation \cite{Wootters1998,Penrose1998}. Entanglement enables us to teleport any quantum state from one point to another, a phenomenon known as quantum teleportation \cite{Bennett1993}. Entanglement is also very useful in quantum cryptography \cite{Yin2020}, quantum imaging \cite{Watts2021}, quantum sensing \cite{Degen2017} and quantum algorithms \cite{Ekert1998}.

To establish long-range entanglement, we typically rely on a classical infrastructure of optical fiber networks \cite{Simon2017}. Entanglement between two nodes connected by a direct optical fiber link can be established by transmitting entangled photons between them \cite{Brito2020}. The entanglement formed by the transmission of a pair of entangled photons is usually nonmaximal due to the presence of various physical losses during the transmission of photons \cite{Brito2020} and imprecise quantum measurements \cite{Schlosshauer2019}. The nonmaximally entangled states formed via an optical fiber are often observed to be in a mixed state due to various sources of noise \cite{Vicente2024}. In addition to optical fibers, long-distance entanglement can also be achieved by transmitting quantum states through satellite based channels \cite{Neumann2018,Zuo2021,Brito2021}. Generally, a network comprising such entangled states — whether maximal or nonmaximal, pure or mixed — is considered a quantum network (QN) \cite{Perseguers2013}.

Establishing maximal entanglement across QNs is essential, as maximally entangled states theoretically enable efficient long-distance quantum communication. However, achieving it is extremely challenging due to the aforementioned losses and noises associated with extant quantum technology. To address this, we first construct a QN where the edges are nonmaximally entangled states. We then focus on converting these nonmaximal states into maximal entangled states to facilitate efficient communication \cite{Perseguers2013}. Various protocols have been proposed to create maximal entanglement across vast distances. These protocols are heavily inspired by the naturally occurring phenomenon of percolation in statistical physics \cite{Perseguers2013}. The successful creation of long-range maximally entangled states using a QN relies heavily on the underlying network structure, the amount of entanglement associated with initial nonmaximally entangled states, and the chosen protocol. Therefore, for a given QN, determining the minimum amount of entanglement required to establish long-range maximal entanglement is a topic of great interest among researchers. While existing protocols provide excellent foundational frameworks, scaling them often incurs high computational and physical resource overheads due to the complexities of finding new and alternate paths and geometric transformations.

\begin{figure}[htbp]
\includegraphics[width=\columnwidth,height=2.8cm]{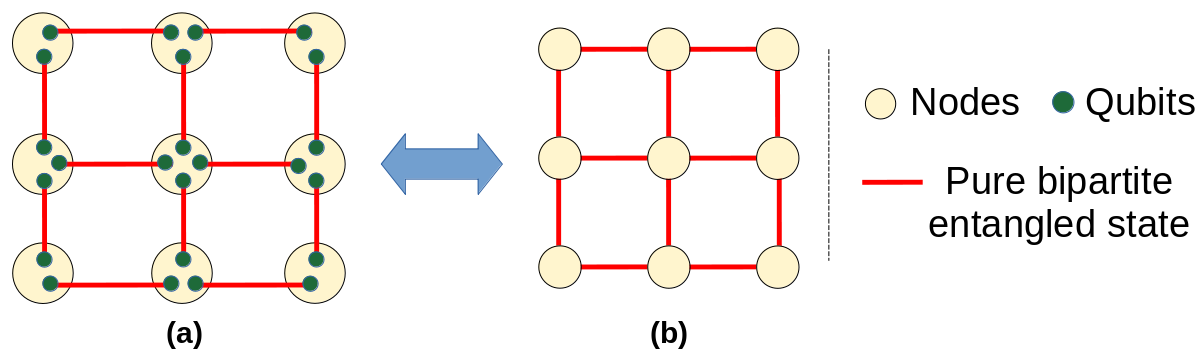}
\caption{(a) $3 \times 3$ square lattice as a quantum network. Each edge represents a pure bipartite entangled state. Initially, all entangled states are nonmaximal. (b) A simpler representation of the quantum network shown in (a). We consistently use the representation from (b) in all our figures.}
\label{fig:qn_fig_1}
\end{figure}

Driven by these factors, our primary motivation Here is to develop a more generalized and resource-efficient framework for successful long-distance quantum communication. Ideally, we could have quantum networks with various topologies. Here, for simplicity we consider two-dimensional square, triangular, and hexagonal lattices where each edge represents a nonmaximally entangled pure state as shown in Fig. \ref{fig:qn_fig_1}. We introduce a protocol, general concurrence percolation (GCP), for establishing maximal entanglement over large distances. GCP achieves densification by amplifying entanglement strictly along shortest paths while maintaining the integrity of all physical nodes. Through GCP, as we show later, we can obtain a network of maximally entangled states from a given network of nonmaximally entangled states at a fundamentally lower percolation threshold. The associated network of maximally entangled states can transmit information between two distant nodes efficiently and securely.

\begin{figure}[htbp]
\includegraphics[width=\columnwidth,height=8cm]{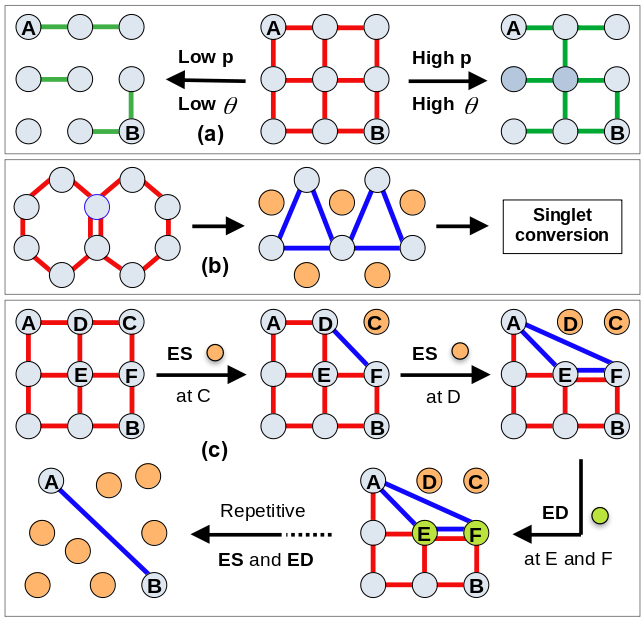}
    
    \caption{(a) Classical entanglement percolation (CEP). nonmaximally entangled states are converted into maximally entangled states (indicated by solid green lines) through singlet conversion. The underlying topology of the network remains unchanged. (b) Quantum entanglement percolation (QEP). Before singlet conversion, entanglement swapping is performed on specific repeater nodes (highlighted in orange) to lower the percolation threshold. During this process, the underlying network topology can change significantly. (c) Concurrence percolation theory (ConPT). The long-range entanglement (indicated by solid blue lines) is established between two distant ($A$ and $B$) nodes through repetitive entanglement swapping (ES) and entanglement distillation (ED).}
    \label{fig:ep_fig_2}
\end{figure}

\section{PRELIMINARIES}
We first recall the protocols that are currently in use to achieve long-range quantum entanglement. These preliminary concepts will be helpful as the  protocol proposed here is based on all these fundamental ideas.
\subsection{Entanglement percolation: Classical and quantum}
\label{sec:perc}
Let ${\cal G}({\cal V},{\cal E})$ denote a QN, where ${\cal V}$ and ${\cal E}$ signify nodes and edges respectively. Here, nodes represent physical stations or locations where quantum information is processed and stored. Edges denote pure entangled states. The number of nodes and edges are denoted by $N_{\cal V}=|{\cal V}|$ and $N_{\cal E}=|{\cal E}|$ respectively. The associated entangled state for a given edge, ${\cal E}_{ij}$, between nodes $i$ and $j$ is expressed as,
\begin{equation}
|\psi_{ij}(\theta)\rangle =\cos\theta |00\rangle +\sin\theta |11\rangle
\label{eq:pure_state}
\end{equation}
For $0<\theta < \pi/4$, we have a nonmaximally entangled pure state. For $\theta = \pi/4$, we have a maximally entangled pure state, while $\theta=0$ indicates an unentangled state. For simplicity, we consider all edges to be identical. We denote the entangled states of every edge, $|\psi_{ij}(\theta)\rangle$, by $|\psi(\theta)\rangle$. In practice, the value of $\theta$ should depend on various factors, such as the transmissivity of the associated channels, the decoherence of entangled photons during transmission, and similar effects \cite{Brito2020}. Due to these physical losses, achieving maximally entangled states (commonly referred to as singlets) is not possible. As mentioned in Sec. \ref{sec:intro}, the resulting entangled states are always mixed states. However, for simplicity, we focus exclusively on nonmaximally entangled pure bipartite states and treat $\theta$ as our variable. We could convert nonmaximal entangled states into maximal ones through singlet conversion for a given $|\psi(\theta)\rangle$. However, the singlet conversion technique is probabilistic and the associated probability of successful conversion is given by
\begin{equation}
p=2\sin^2\theta.
\label{eq:SCP}
\end{equation}
The nonmaximally entanglement state will be completely lost in the event of an unsuccessful conversion. It may appear that for a given quantum state entanglement is increasing.  However, using local operation and classical communication (LOCC) alone will not allow us to increase or generate entanglement. Because of this, we are limited to creating singlet states probabilistically, which obey the conservation of entanglement. This singlet conversion protocol associated with a QN in the thermodynamic limit is often referred to as classical entanglement percolation (CEP) due to its conceptual correlation with the phenomenon of classical bond percolation in networks \cite{Perseguers2013}.  For two distant nodes, the existence of at least one path of singlets allows us to establish long-distance quantum communication through entanglement swapping \cite{Ekert1993,Bose1999}. The associated percolation threshold value in terms of $\theta$, namely $\theta_{T}$, provides us with the minimum value of $\theta$ for which a giant connected component of maximally entangled states arises. Before and after singlet conversion, the topology of the QN is identical to that of the underlying infrastructure, as shown in Fig. \ref{fig:ep_fig_2}(a).

In CEP, the topology of a QN does not change. However, in another protocol, known as quantum entanglement percolation (QEP), the topology of the QN is modified before singlet conversion, as shown in Fig. \ref{fig:ep_fig_2}(b). Usually, this alteration in topology occurs due to entanglement swapping at some specific nodes, which act as quantum repeaters \cite{Shchukin2022}. By targeting these nodes for entanglement swapping, we change the overall topology of the QN to lower the value of the percolation threshold  \cite{Acin2007}. However, during this process, the repeater nodes can become disconnected from the QN of singlet states.  Thus, the choice of repeater nodes is somewhat arbitrary, and they can become disconnected after entanglement swapping. Further, in most cases, QEP cannot decrease the percolation threshold.

\subsection{Concurrence percolation theory}
\label{sec:ConPT}

Creating at least one path of multiple maximally entangled states between a source and a remote destination is our primary motivation in CEP and QEP protocols. Nielson's majorization criterion provides a singlet-conversion method, which is probabilistic. Consequently, most nonmaximally entangled states can be lost during the singlet conversion when $\frac{\pi}{4} \gg \theta$, even for  $\theta > \theta_{T}$. Therefore, in concurrence percolation theory (ConPT), we emphasize ``sponge-crossing'' paths rather than generating a giant cluster of maximally entangled states \cite{Meng2021}. We could employ deterministic LOCC to establish a direct link of a nonmaximally entangled state between two distant nodes, as shown in Fig. \ref{fig:ep_fig_2}(c). This direct link allows us to communicate quantum information upon singlet conversion. ConPT is conceptually similar to CEP because both are based on percolation. However, in ConPT, instead of probability, we rely on concurrence, $c$, an important measure of entanglement. For $|\psi (\theta)\rangle$, we have
\begin{equation}
c=\sin2\theta
\label{eq:concurrence}
\end{equation}

In addition to entanglement swapping, we also apply entanglement distillation. Effectively, this technique allows us to concentrate our resources, sacrificing a larger volume of nonmaximal states to generate a smaller, yet significantly more pristine collection of purified entangled pairs.
By using repetitive entanglement swapping (series rule) and entanglement distillation (parallel rule), we can use the star-mesh transform technique to obtain an entangled state between a distant source and the destination \cite{Meng2021}. The associated connectivity rules for both CEP and ConPT can be expressed as series and parallel rules, as shown in Table II of \cite{Meng2021}.  It has been reported \cite{Meng2021} that the associated percolation threshold for ConPT, in terms of $\theta$, is the lowest compared to those for CEP and QEP.

\section{General concurrence percolation}
\label{sec:GCP}
Our method combines the three entanglement percolation protocols, namely, CEP, QEP, and ConPT, and offers a general protocol for establishing quantum communication, taking into account the maximal entanglement between two remote nodes.

\begin{figure}[htbp]
%\centering
\includegraphics[width=\columnwidth, height=\columnwidth]{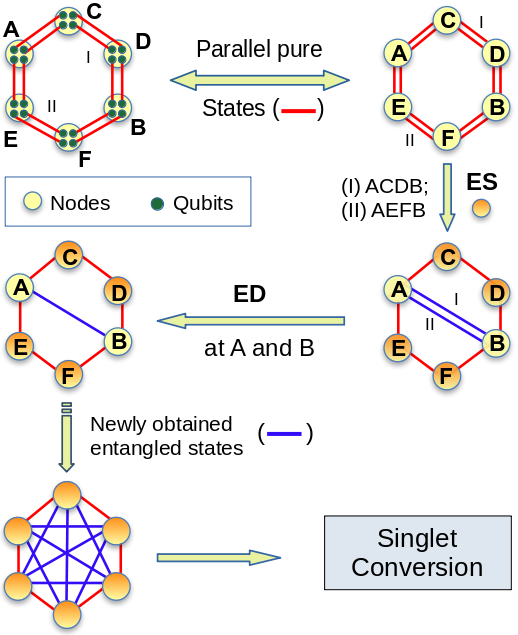}
%\caption{General Concurrence Percolation (GCP). Initially, there are parallel edges connecting two neighboring nodes, which can be interpreted as being linked by a state composed of two identical copies of a two-qubit nonmaximally entangled state (\red{\bf{\LARGE \textemdash}}). Alternately, $| \psi^{\prime} \rangle = | \psi^{\otimes 2} \rangle$, as shown in Fig. 3 of \cite{Acin2007}. First, entanglement swapping (ES) is performed along the paths: (I) $ACDB$, and, (II) $AEFB$. Subsequently, entanglement distillation ($ED$) is performed at $A$, and, $B$ to create a new entangled state (\magenta{\bf{\LARGE \textemdash}}) between them. This process of ES and ED is iterated across the network to generate entangled states (\magenta{\bf{ \LARGE \textemdash}}) between all other node pairs.}
\caption{General concurrence percolation (GCP). Initially, there are parallel edges connecting two neighboring nodes, which can be interpreted as being linked by a state composed of two identical copies of a two-qubit nonmaximally entangled state (indicated by solid red lines). Alternately, $| \psi^{\prime} \rangle = | \psi^{\otimes 2} \rangle$, as shown in Fig. 3 of \cite{Acin2007}. First, entanglement swapping (ES) is performed along the paths: (I) $ACDB$ and (II) $AEFB$. Subsequently, entanglement distillation (ED) is performed at $A$ and $B$ to create a new entangled state (indicated by solid blue lines) between them. This process of ES and ED is iterated across the network to generate new entangled states between all other node pairs.}
\label{fig:GCP}
\end{figure}

Node pairs in a network possessing a shortest path between them that is comparable to the diameter of the network, can be referred to as distant nodes. Obviously, a good number of edges must be traversed to move from one end node to the other end node of a distant node pair. Figure \ref{fig:GCP} presents a small network that illustrates how we can navigate this structural issue. Even when two nodes lack a direct link, we can bridge the gap by exploiting the quantum capacity for multiple parallel entangled states between node pairs \cite{Acin2007}.
Consider the specific case of nodes $A$ and $B$ in the diagram. Although no direct edge connects them, we can have nonmaximal entangled states between neighboring nodes. To establish a connection, we apply the ``series rule'' of entanglement swapping along two distinct pathways: (I) $ACDB$ and (II) $AEFB$. This operation effectively generates two parallel entangled links between $A$ and $B$. We can then refine these connections using the ``parallel rule,'' as detailed in Table II of \cite{Meng2021}, which allows us to distil these multiple instances into a single entangled state.  Ultimately, this methodology empowers us to engineer long-range connections between remote nodes, successfully overcoming the structural limitations of the underlying network structure.

However, the value of $\theta$ associated with these newly established entangled states would differ and depend on the number and length of the shortest paths between the related sources and destinations. Therefore, the corresponding singlet conversion probability, as shown in Eq. \ref{eq:SCP}, would also differ for different entangled states. This protocol is referred to as general concurrence percolation (GCP). Only the shortest paths of the network have been taken into consideration to establish entanglement between two distant nodes in a large network \cite{Malik2022}. In GCP, the associated variable is considered to be either $c$ or $\theta$.

\section{Methods}
\label{sec:method}
For a given network of nonmaximally entangled states, we compute the length $l_{ij}$ and number $n_{ij}$ of all feasible shortest paths between each randomly selected pair of nodes, $i$ and $j$. Here for notational simplicity, for a predecided $i$ and $j$, let us temporarily refer to $l_{ij}$ and $n_{ij}$ as $l$ and $n$, respectively. In Fig. \ref{fig:GCP}, we observe that there are two shortest paths between $A$ and $B$ with length $3$, i.e., $l=3$ and $n=2$. For a given pair of nodes, $l=1$ and $n=1$ signify a direct  link between $i$ and $j$. A nonmaximally entangled state $|\psi (\theta)\rangle$ can be achieved between $i$ and $j$, as shown in Eq. \ref{eq:pure_state}. $|\psi (\theta)\rangle$ can be converted into singlets with the probability given in Eq. \ref{eq:SCP} through CEP. For $l=2$ and $n=1$, we have to perform entanglement swapping through QEP. For two entangled states with concurrence $c_1$ and $c_2$, the associated concurrence, $c^\prime$, of the new entangled state after entanglement swapping following the series rule will be $c^\prime=c_1c_2$ \cite{Meng2021}. For a given state $|\psi (\theta)\rangle$, the value of the concurrence $c$ can be calculated using Eq. \ref{eq:concurrence}. For any arbitrary value of $l$, we have
\begin{equation}
c^\prime=\prod_{i=1}^{l}c_i.
\label{eq:series_rule_c}
\end{equation}
For simplicity, we consider all initial nonmaximally entangled states to be identical. Therefore, the values of $\theta$ and $c$ associated with every initial entangled state $|\psi (\theta)\rangle$ are identical, i.e., $c_1=c_2=c$. From Eq. \ref{eq:series_rule_c}, we have $c^\prime=c^2$. Similarly, for paths with length $l$, we have $c^\prime=c^{l}$.

For shortest paths with $l>1$ and $n>1$, we must first carry out entanglement swapping for each shortest path. Then, we may execute entanglement distillation on the $n$ number of new entangled states between $i$ and $j$. Following the process of entanglement distillation, the new concurrence value of the new entangled state following the parallel rule \cite{Meng2021} is,
\begin{multline}
\frac{1+\sqrt{1-(c^\prime)^2}}{2} \, = \, max\{\frac{1}{2} ,\,  \prod_{i=1}^{n}\frac{1+\sqrt{1-(c_i)^2}}{2}\}.
\label{eq:parallel_rule_c}
\end{multline}
Using this repetitive swapping [Eq. \ref{eq:series_rule_c}] and distillation [Eq. \ref{eq:parallel_rule_c}], we establish a direct entanglement link between two distant nodes, similar to ConPT.  Unlike ConPT, here, we consider only the shortest paths between two nodes. After establishing entanglement between all possible pairs of nodes, we perform singlet conversion using Eq. \ref{eq:SCP}. However, the singlet conversion probability for each edge is different and depends on associated values of $c$ and $\theta$. From Eqs. \ref{eq:series_rule_c} and \ref{eq:parallel_rule_c}, we obtain the values of concurrence $c$ associated with new entangled states. The associated values of $\theta$ can be easily determined by a change in variables as described in Eq. \ref{eq:concurrence}. In the case of entanglement swapping, the new entangled state is a mixed state. We have four pure states $c^{\prime}_k$ with probability $\omega_k$. The concurrence of the new mixed entangled state is $C_k=\sum_{k=1}^4 \omega_k c^{\prime}_k$. However, it has been reported that, if we consider $XZ$ bell state for projection all $c^{\prime}_k$ will be identical with equal probability $\omega_k$ \cite{Meng2021,Perseguers2008}. Therefore, we would have $C_k= c^{\prime}_k$. After obtaining a direct link of nonmaximally entangled states, we apply the singlet conversion technique to get a maximally entangled state.

Upon successful singlet conversion, we will have a network of maximally entangled states. We calculate the fraction of the largest connected component, also known as the giant component, $f_{gc}$, for different values of $\theta$. If $N_{gc}$ is the number of nodes in the largest connected component for a particular value of $\theta$, then the associated fraction of the giant component will be $f_{gc}=N_{gc}/N$. The total number of nodes in the QN is indicated here by $N \equiv N_{\cal V}$. The value of $\theta$ that gives rise to a giant component is known as the percolation threshold value, or $\theta_{th}$. We should observe a continuous transition of $f_{gc}$ with $\theta$ for finite-size lattices. Conventionally,  the values of $\theta$ associated with $f_{gc}=0.5$ are recognized as $\theta_{th}$ in the case of finite-size networks. However, we must perform finite-size scaling to calculate the percolation threshold value for $N \to \infty$. Here, the percolation threshold for infinite lattices associated with the GCP protocol is denoted as $\theta_{T}=\theta_{th}|_{N \to \infty}$. In the case of $\theta<\theta_{T}$, the giant component is absent; that is, the associated $f_{gc}$ of the largest connected component is very small. Therefore, for $\theta<\theta_{T}$, entanglement can not percolate through that network configuration, hence preventing the achievement of long-distance quantum communication. However, we will have a giant spanning cluster of maximally entangled states for $\theta>\theta_{T}$. Therefore, long-distance quantum communication may be feasible for $\theta>\theta_{T}$ in a given QN. Similarly, we can also have $c_{th}$ and $c_T$, which can be easily obtained under a change in variables from $\theta$, as shown in Eq. \ref{eq:concurrence}.

\begin{figure}
      \centering
	    % \begin{subfigure}{\linewidth}
		\includegraphics[width=\linewidth, height=5cm]{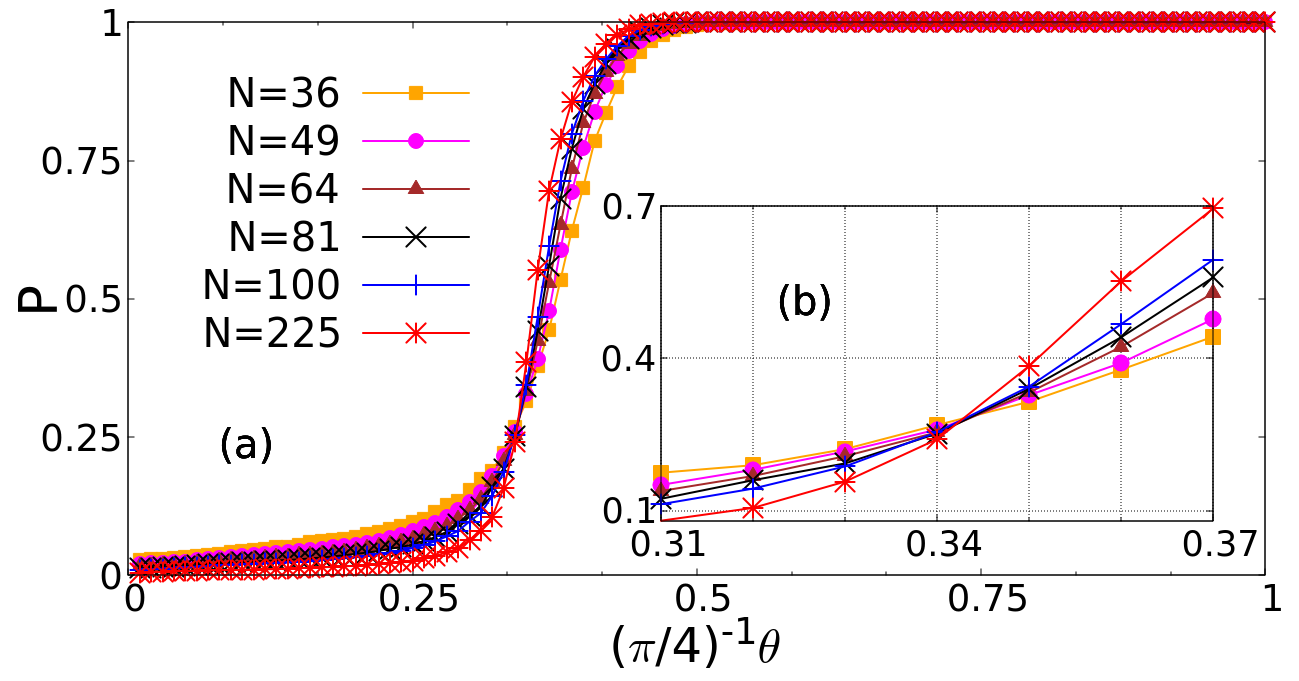}
		\caption{(a) Density of the spanning cluster $P$ versus $(\pi/4)^{-1}\theta$ for various finite-sizes of square lattice, i.e., $N=36, 49, 64, 81, 100, 225$. (b) The intersection of all curves for various $N$ decides the value of the percolation threshold, $\theta_{T}$. Here, $E_N=1000$ ensembles.}
		\label{fig:threshold_square}
	   %\end{subfigure}
\end{figure}
%	\vfill
\begin{figure}
      \centering
%	     \begin{subfigure}{\linewidth}
		 \includegraphics[width=\linewidth, height=5cm]{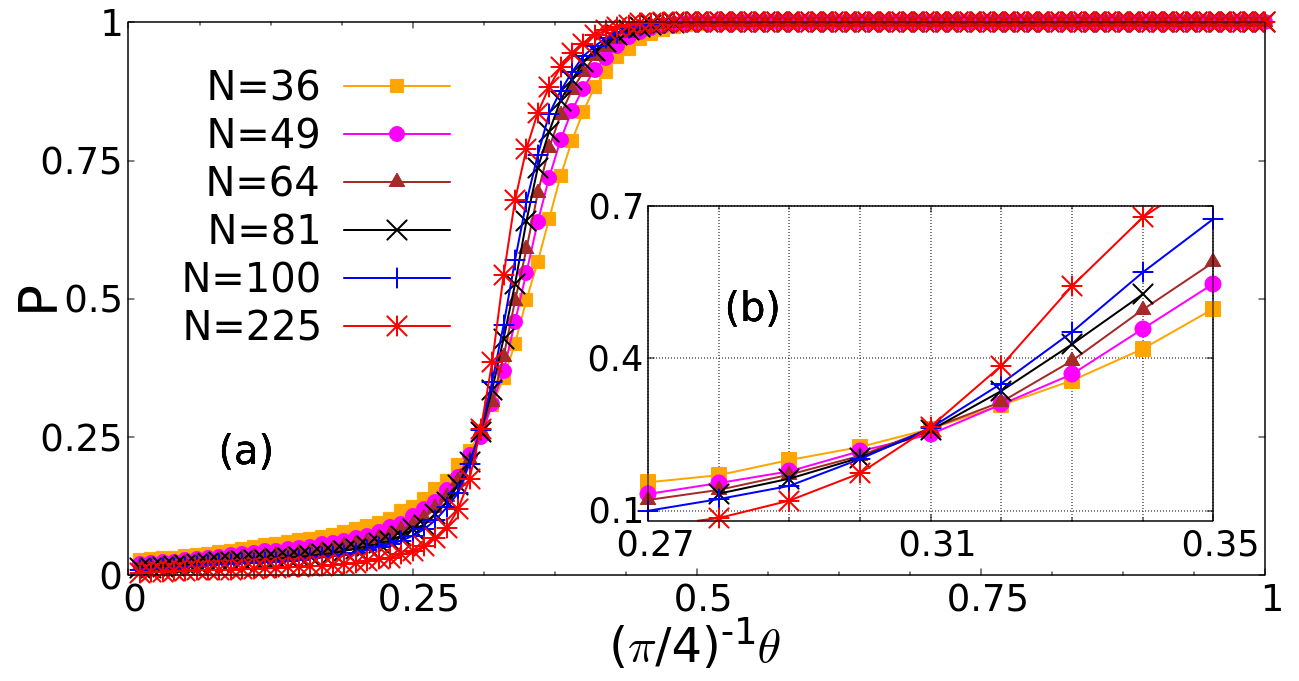}
		 \caption{(a) Density of the spanning cluster $P$ versus $(\pi/4)^{-1}\theta$ for various finite-sizes of triangular lattice, i.e., $N=36, 49, 64, 81, 100, 225$. (b) The value of the percolation threshold, $\theta_T$, is determined by the intersection of all curves for various $N$. Here, $E_N=1000$ ensembles.}
		 \label{fig:threshold_tri}
%	      \end{subfigure}
\end{figure}
%	\vfill
\begin{figure}
      \centering
%	     \begin{subfigure}{\linewidth}
		 \includegraphics[width=\linewidth, height=5cm]{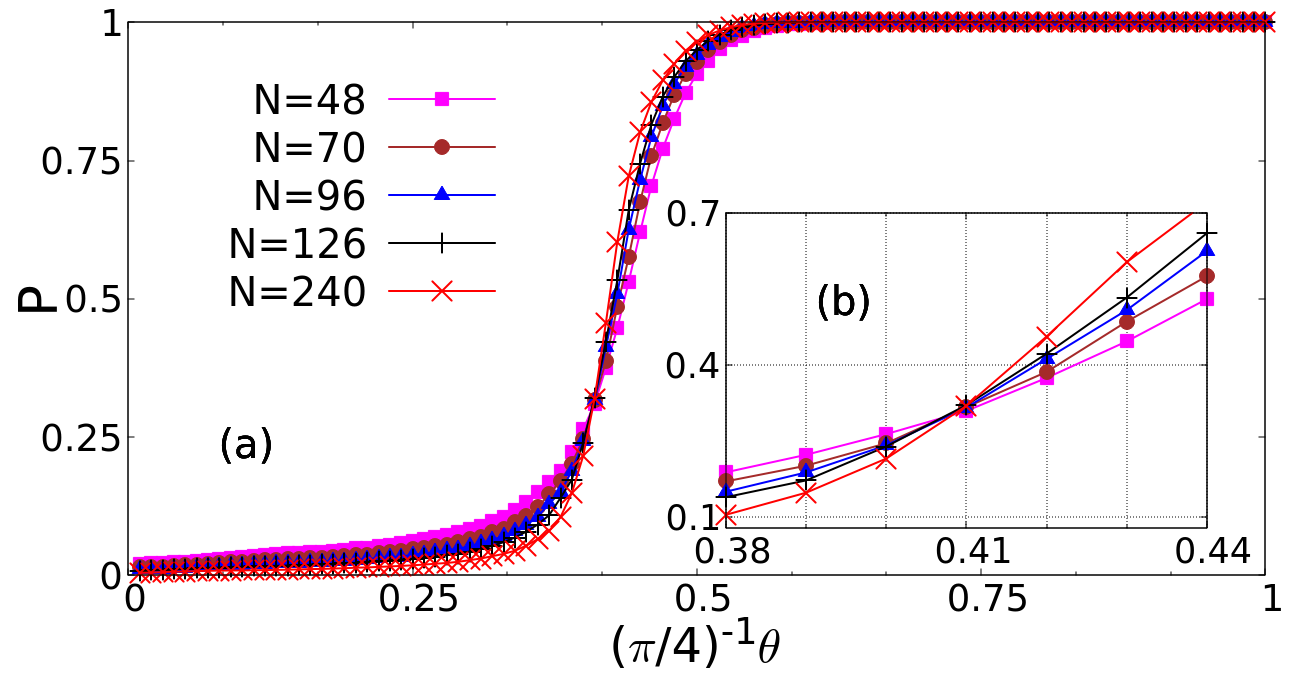}
		 \caption{(a) Density of the spanning cluster $P$ versus $(\pi/4)^{-1}\theta$ for various finite-sizes of hexagonal lattice, i.e., $N=48, 70, 96, 126, 240$. (b) The intersection of all curves for various $N$ renders the value of the percolation threshold, $\theta_T$. Here, $E_N=1000$ ensembles.}
		 \label{fig:threshold_hexa}
%	      \end{subfigure}
\end{figure}

In percolation theory, the system is characterised by the order parameter $P(c,L)$, defined for a finite $L \times L$ lattice as the density of the largest (spanning) cluster, i.e., $P(c,L) = s_{\max}/L^2$. Here, $s_{max}$ is the number of nodes (sites) in the largest connected component. On the other hand, we define $f_{gc} = N_{gc}/N$ as the fraction of nodes in the largest connected component of the network. Although the quantities $P (c, L)$ and $f_{gc}$ seem more relevant in the context of percolation and networks respectively, they are numerically identical in our setup. That is because, for the lattices considered here, $N = L^2$ and the spanning cluster is identified with the largest connected component, i.e., $s_{max}=N_{gc}$. Hence, in the following, we use $P$ and $f_{gc}$ interchangeably. In particular, $P(c,L)$ is computed numerically as the fraction of nodes in the largest connected component. Therefore,

\begin{equation}
P(c,L) = \frac{s_{\max}}{L^2} = \frac{N_{gc}}{N} \equiv f_{gc}.
\label{eq:P_f_gc}
\end{equation}

We can write the finite-size scaling behavior of the spanning cluster density as
\begin{equation}
P=L^{-\beta/\nu}f(L/\xi)
\label{eq:P_1}
\end{equation}
Here, $\xi$ represents the correlation length; $\xi \propto \xi_0|c-c_{th}|^{-\nu}$ and $\nu$ is the associated critical exponent. $\beta$ is the critical exponent associated with $P$, such that $P(c)\propto |c-c_{th}|^\beta$. From percolation theory, using Eq. \ref{eq:P_1}, we have the scaling form of $P$ as
\begin{equation}
PN^{\beta/d\nu}=f(N^{1/d\nu}(c-c_{th}))
\label{eq:P_2}
\end{equation}
Here, $N=L^d$ and for $2D$ lattices, d=2. To test the scaling anastz we plot $y=PN^{\beta/d\nu}$ against $x=(c-c_{th})N^{1/d\nu}$. Further, we explore whether the data for various system sizes collapse onto the same curve for particular value of $\beta$ and $\nu$ \cite{Stauffer1994}. We can also calculate the value of $\theta_T$ using the scaling ansatz \cite{malthe2024percolation, Gimenez2025},

\begin{equation}
(\theta_{th}-\theta_{T}) \propto N^{-1/2\nu}
\label{eq:theta}
\end{equation}

The value of $\theta_T$ is obtained via finite-size scaling using Eq. \ref{eq:theta}, by extrapolating $\theta$ as a function of $N^{-1/(2\nu)}$ to the limit $N^{-1/(2\nu)} \to 0$, which corresponds to $N \to \infty$. The intercept of the linear least-squares fit in this limit also helps us to determine $\theta_T$. In our analyses, we use the critical exponent $\nu$, which is determined independently from the data-collapse, given by Eq. \ref{eq:P_2}.

In our simulations, we consider three kinds of 2D lattices as QNs, namely, square, triangular and hexagonal. Here, the edges of these lattices represent nonmaximally entangled pure states between two neighbors, i.e., $l=1$. However, for non-neighboring nodes, we can determine the number, $n$, and length, $l$, of the shortest paths. We estimate the values of $n$ and $l$ associated with a pair of nodes using Dijkstra's algorithm \cite{Dijkstra1959} for unweighted graphs. However, determining $n$ and $l$ for all node pairs is computationally non-trivial. A simple method for calculating $n$ and $l$ in the case of a square lattice \cite{Malik2022} was provided to restrict undue expenditure of computational resources. In Appendix, we calculate $n$ and $l$ for any random pair of nodes in triangular lattices. Using the existing infrastructure of QN, we establish direct entangled states between all non-neighboring node pairs through repetitive entanglement swapping [Eq. \ref{eq:series_rule_c}] and distillation [Eq. \ref{eq:parallel_rule_c}]. Thus, we obtain a fully connected network of nonmaximally entangled states. Here, the value of $\theta$ and $c$ associated with each new entangled state will be different. After that, we perform singlet conversion according to Eq. \ref{eq:SCP}. We calculate $P$ for various values of $\theta$ and calculate the value of the percolation threshold $\theta_T$ associated with each lattice. The algorithm we follow in our numerical simulations is detailed in Appendix \ref{sec:appendix_B}. The code used in this work is available in Ref.~\cite{gcp_code}.

\section{Results and Discussion}
\label{sec:results}

\begin{figure}
      \centering
		\includegraphics[width=\linewidth, height=5cm]{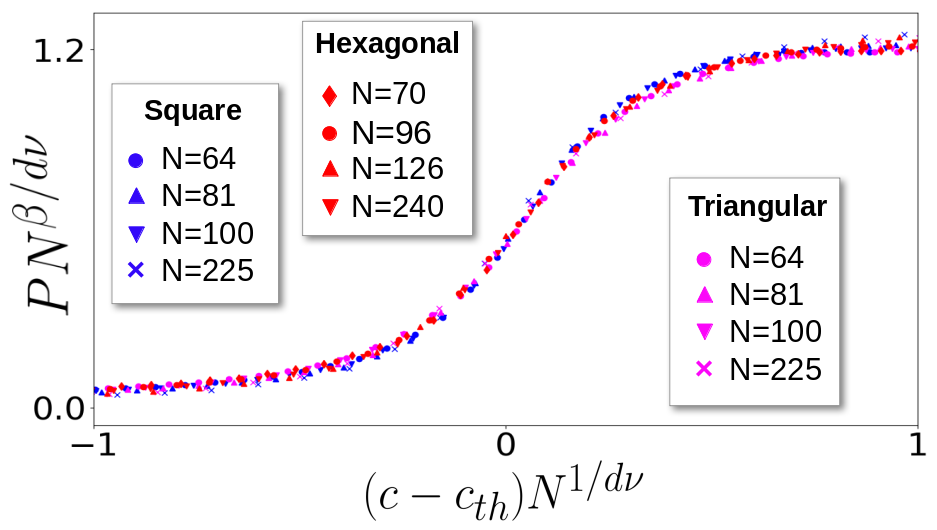}
		\caption{Data collapse for the density of the spanning cluster, $P(c,N) \equiv P$. The finite-size analysis for 2D lattices ($d=2$), including square, triangular, and hexagonal lattices, is shown in Figs. \ref{fig:threshold_square}, \ref{fig:threshold_tri}, and \ref{fig:threshold_hexa}, respectively. Here, finite-size scaling has been performed with $\nu=1.3(4)$ and $\beta=0.11(5)$. Results are for $E_N=1000$ ensembles.}
		\label{fig:dc_c}
\end{figure}

In Figs. \ref{fig:threshold_square}(a), \ref{fig:threshold_tri}(a), and \ref{fig:threshold_hexa}(a), we observe the variation of $P$ with respect to $(\frac{\pi}{4})^{-1}\theta$, for square, triangular and hexagonal lattice respectively at various system sizes. Here, we observe a continuous phase transition for finite-size lattices. In the case of $N \to \infty$, we have multiple disconnected components of small clusters for $\theta < \theta_{T}$. For $\theta > \theta_{T}$, we have a spanning cluster, i.e., the largest connected (giant) component. For a given lattice,  the intersection point of the curves corresponding to $P(c,N)$ for different finite sizes, i.e., different $N$, is typically considered to be $\theta_T$ \cite{Binder1992, Stauffer1994,Meng2021}. In Figs. \ref{fig:threshold_square}(b), \ref{fig:threshold_tri}(b), and \ref{fig:threshold_hexa}(b), we observe the intersection points of the curves corresponding to finite-size lattices. However, to accurately determine the value of $\theta_T$, it is necessary to perform a finite-size scaling analysis and evaluate the critical exponents.

\begin{figure*}
      \centering
		\includegraphics[width=\textwidth, height=5.5cm]{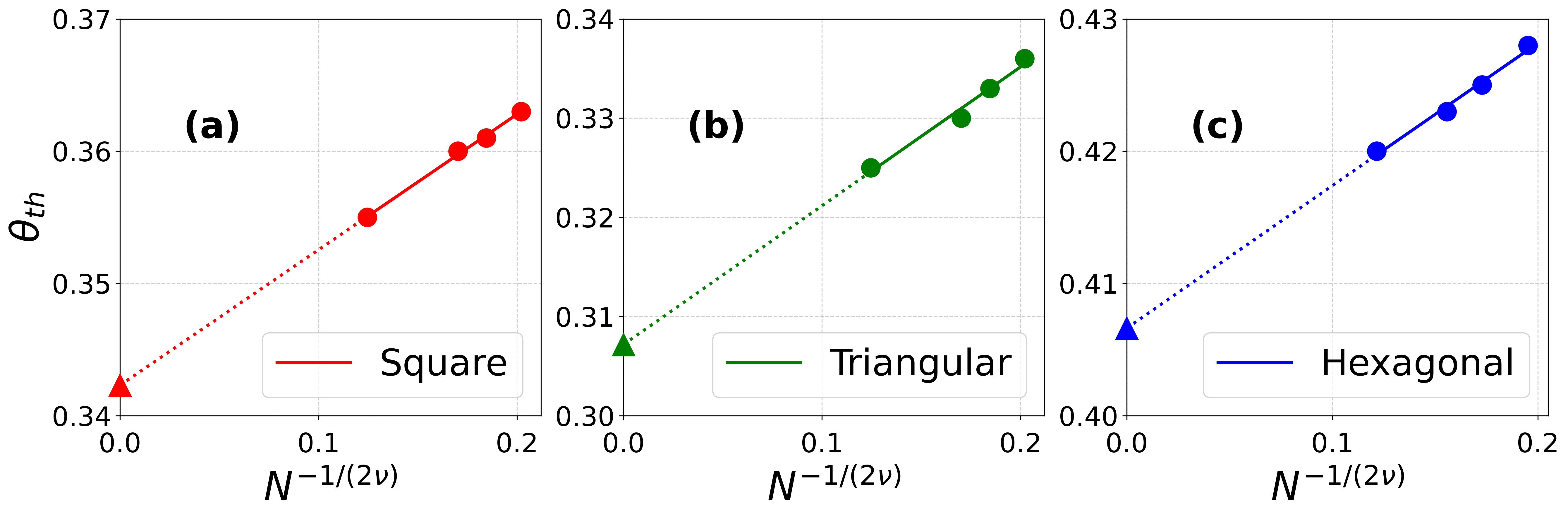}
		\caption{Finite-size scaling of $\theta_{th}$ {\it versus} $N^{-1/(2\nu)}$ for the: (a) square ($N = 64, 81, 100, 225$), (b) triangular ($N = 64, 81, 100, 225$), and (c) hexagonal ($N = 70, 96, 126, 240$) lattice. The solid and dashed lines respectively represent a linear least-square fit and its extrapolation. The intercept at $N \to \infty$ yields the percolation threshold in the thermodynamic limit, $\theta_{T}$, using $\nu = 1.3(4)$ as obtained in Fig.~\ref{fig:dc_c}. The resulting values of $\theta_T$ [in units of $(\pi/4)^{-1}\theta$] for the square, triangular, and hexagonal lattice are $0.342(5)$, $0.307(7)$, and $0.407(5)$ respectively.}
		\label{fig:theta_fit}
\end{figure*}

Characterising phase transitions by identifying the associated critical exponents is typically of greater interest. We can determine which universality class a process belongs to by looking at the values of the critical exponents and the relationship between them. In Fig. \ref{fig:dc_c}, we perform data collapse for $P(c,N) \equiv P$ using Eq. \ref{eq:P_2}. We observe that the data points for various system sizes collapse onto the same curve for $\nu=1.3(4)$ and $\beta=0.11(5)$. It is well-established that for percolation on a 2D lattice, the values of the critical exponents are $\nu = \frac{4}{3} \approx 1.33$ and $\beta = \frac{5}{36} \approx 0.1389$ \cite{Stauffer1994}. Therefore, it is reasonable to conclude that our protocol, GCP, belongs to the same universality class.  It has been reported that CEP and QEP belong to the percolation universality class \cite{Meng2021}. However, ConPT belongs to a different universality class \cite{Hu2025}. The value of $\nu$, thus obtained, is utilized to calculate $\theta_T$ through finite-size scaling in Fig.~\ref{fig:theta_fit}. The value of $\theta_T$ is calculated for the square, triangular, and hexagonal lattice in Figs.~\ref{fig:theta_fit}(a), \ref{fig:theta_fit}(b), and \ref{fig:theta_fit}(c) respectively using the scaling ansatz. It can be verified that the values obtained in Fig.~\ref{fig:theta_fit} are in agreement with those obtained from Figs \ref{fig:threshold_square}, \ref{fig:threshold_tri} and \ref{fig:threshold_hexa}. We compare these values of $\theta_T$ with other available protocols in Table \ref{tab:per_thres}. We observe that the GCP protocol provides the lowest values of percolation threshold compared to those of other protocols.

\begin{table}[t]
\centering
\caption{The threshold values, $\theta_{T}$ [in units of $(\pi/4)^{-1}\theta$], associated with GCP protocol in the case of square, triangular and hexagonal lattices are calculated from Fig. \ref{fig:theta_fit}. GCP provides the lowest threshold value in comparison to other protocols. Results are for $E_N=1000$ ensembles.}
\label{tab:per_thres}
\begin{tabular}{p{2.5cm} p{1.8cm} p{1.8cm} p{1.8cm}} 
\toprule
Protocol & Square & Triangular & Hexagonal  \\
\midrule
CEP \cite{Acin2007} & 0.670 & 0.545 & 0.777 \\
QEP \cite{Acin2007, Cuquet2009, Perseguers2008} & 0.670 & 0.545 & 0.761 \\
QEP-GHZ \cite{Perseguers2010} & 0.584 & 0.481 & 0.745 \\
ConPT \cite{Meng2021} & 0.42(8) & 0.32(8) & 0.51(8) \\
\textbf{\emph{GCP}} & 0.342(5) & 0.307(7) & 0.407(5) \\
\bottomrule
\end{tabular}
\end{table}

\begin{figure}[htbp]
%\centering
\includegraphics[width=\columnwidth, height=5cm]{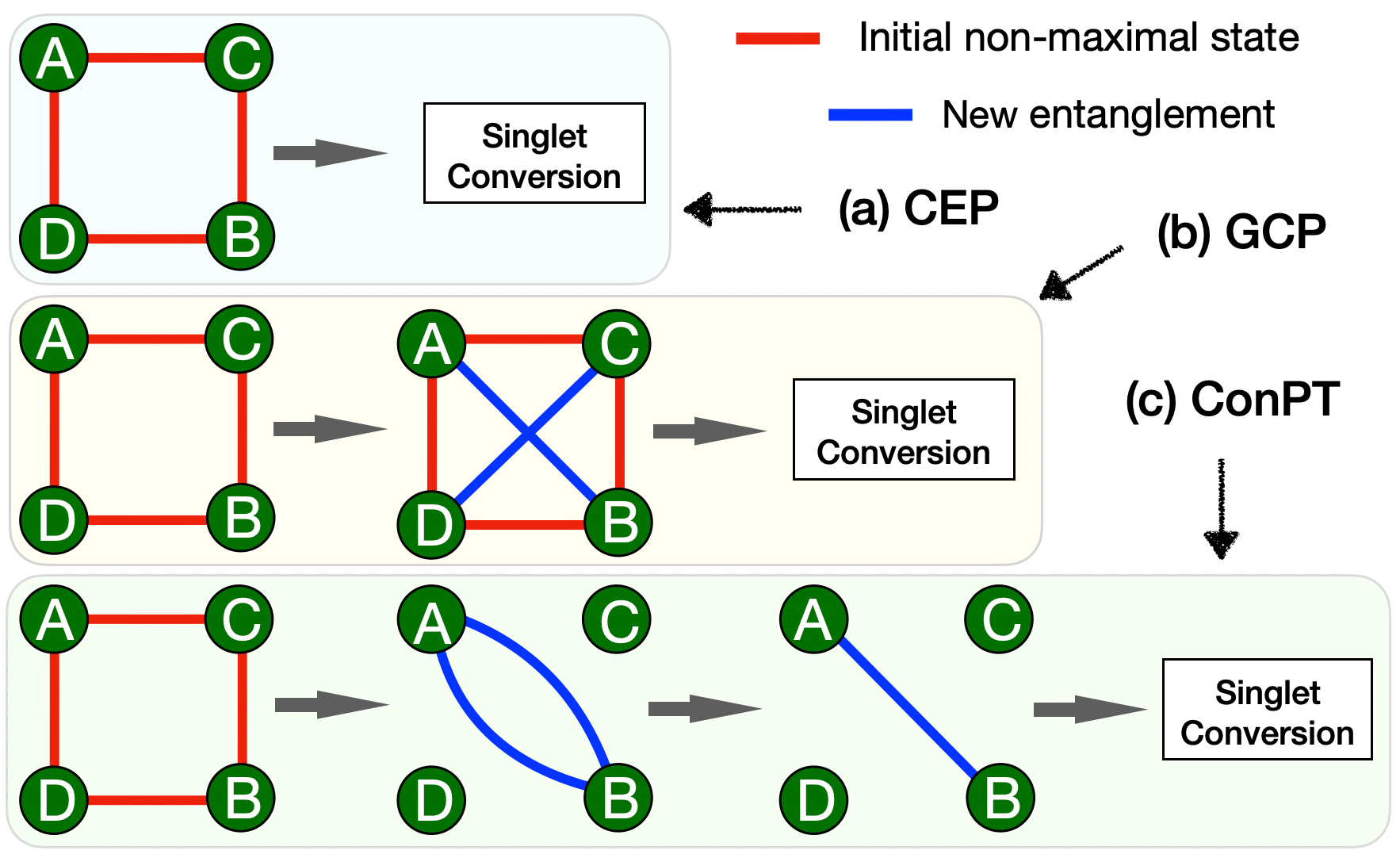}
\caption{Comparison of entanglement percolation protocols on a $2 \times 2$ plaquette. (a) CEP: Direct probabilistic singlet conversion on the initial sparse lattice. (b) GCP (densification): Shortest-path LOCC establishes virtual edges [$(A,B)$, $(C,D)$], preserving all nodes and transforming the sparse lattice into a fully connected complete graph prior to conversion. (c) ConPT (reduction): The star-mesh transform physically consumes intermediate nodes ($C$ and $D$) via LOCC, collapsing the network into a single effective edge.}
\label{fig:comp}
\end{figure}

In comparison with CEP and QEP, ConPT and GCP provide a much lower percolation threshold. Although ConPT and GCP incorporate the quantum advantage, they address different aspects of percolation. To analytically demonstrate the fundamental advantage of GCP over existing protocols, we consider a minimal model in the form of a $2 \times 2$ square plaquette as shown in Fig. \ref{fig:comp}. We aim to establish maximal entanglement between two unconnected diagonal nodes, $A$ and $B$, connected via intermediate nodes $C$ and $D$. Each of the four edges initially holds a nonmaximally entangled pure state characterised by $\theta$, corresponding to a singlet conversion probability, $p$, and concurrence, $c$.

In CEP, singlet conversion is attempted directly on the bare edges. Here, we have two shortest paths of length $l=2$ ($A \to C \to B$ and $A \to D \to B$). Let $p_i$ denote the probability associated with $i^{th}$ path. Using the series and parallel rules of CEP \cite{Meng2021}, we obtain the probability of successfully establishing at least one maximal connection between $A$ and $B$ as,
\begin{equation}
p_{{}_{\text{CEP}}} = 1 - \prod_{i=1}^{2} (1-p_i) = 1 - (1-p^2)^2 = 2p^2 - p^4.
\label{eq:p_cep}
\end{equation}

Under ConPT, the network is reduced via the star-mesh transform. For the given plaquette, this equates to entanglement swapping [Eq. \ref{eq:series_rule_c}] along the two shortest paths yielding an effective concurrence $c^\prime = c^2$, followed by entanglement distillation [Eq. \ref{eq:parallel_rule_c}] to the distillation of the shortest paths into a single effective edge. The new effective concurrence, $c_{\text{new}}$, satisfies:
\begin{equation}
\frac{1+\sqrt{1-c_{\text{new}}^2}}{2} =\, max \{\frac{1}{2} ,\, \left( \frac{1+\sqrt{1-(c^2)^2}}{2} \right)^2\}.
\label{eq:toy_parallel}
\end{equation}
Solving for the successful singlet conversion probability of this deterministically amplified state yields:
\begin{equation}
p_{{}_{\text{ConPT}}}  = 1 - \sqrt{1-c_{\text{new}}^2}.
\label{eq:p_conpt}
\end{equation}

GCP, conversely, does not reduce the network to a single edge. Instead, it deterministically amplifies entanglement via LOCC along the shortest paths to draw new edges, transforming the sparse plaquette into a fully connected complete graph while retaining all perimeter nodes. The new diagonal edges $(A,B)$ and $(C,D)$ each attain the probability $p_{\text{AB}} = p_{\text{CD}} = 1 + c^4/2 - \sqrt{1-c^4}$. Because multiple dependent pathways now exist (the direct edge, perimeter paths, and cross-paths via the newly formed edge $(C,D)$), we have the probability of having at least one maximal connection between $A$ and $B$:
\begin{equation}
p_{{}_{\text{GCP}}}  = 1 - \prod_{i=1}^{4} (1-p_i) = 1 - ((1-p^2)^2 (1-p_{AB}) (1-p^2 p_{CD})).
\label{eq:p_gcp}
\end{equation}

GCP mathematically guarantees a higher localized probability of establishing a singlet. For an initial state of $\theta = \pi/8$ (associated with $c \approx 0.707$, $p \approx 0.293$), $p_{{}_{\text{CEP}}} \approx 0.164$ and $p_{{}_{\text{ConPT}}} \approx 0.259$, whereas GCP achieves an amplified probability of $p_{{}_{\text{GCP}}} \approx 0.395$.

While the amplification is significant even at the level of the small network exemplified in our minimal model, the real advantage of GCP emerges when scaled to macroscopic networks. The core diﬀerence between GCP and ConPT lies in their difference of approach in geometry and physical resource management. The following aspects deserve mention.

ConPT iteratively disconnects intermediate nodes to collapse the network into a single effective edge, as shown in Fig. \ref{fig:comp}. This reduction is achieved through the repetitive application of the Star-Mesh transform, which maps a star graph of $n$ nodes to a complete graph of $(n-1)$ nodes. Conversely, GCP preserves the underlying network nodes while establishing direct edges between all unconnected node pairs. By introducing these additional links, GCP transforms the initial sparse lattice into a dense, fully connected complete graph.

By performing its final classical percolation on a far denser graph rather than the original sparse lattice, GCP achieves a fundamentally lower global percolation threshold compared to ConPT, as shown in Table \ref{tab:per_thres}.

In ConPT, we focus on establishing a single path between a source, $s$, and destination, $t$, in the form of a sponge-crossing probability \cite{Meng2021,Malik2022}. In doing so, we consider all possible paths between $s$ and $t$, as shown in Fig. \ref{fig:ep_fig_2}(c). However, for large networks, we could consider only the shortest and second-shortest paths \cite{Malik2022}. In contrast to ConPT, the present protocol (GCP) deals with only the shortest paths. In GCP, if required, we could incorporate all other paths. However, that would not be useful because we can achieve a lower percolation threshold simply by considering only the shortest paths, as shown in Table \ref{tab:per_thres}.  On the other hand, GCP does not require the specification of $s$ and $t$. Instead, we need to create only a spanning cluster of maximally entangled states.

A recent study demonstrated that ConPT falls into a distinct class, primarily due to the inclusion of non-shortest paths \cite{Hu2025}. In contrast, GCP focuses exclusively on shortest paths. This seems to be the prime underlying reason that GCP belongs to the standard percolation universality class. Therefore, the fundamental insight of GCP is that we can achieve lower percolation threshold, while simultaneously benefiting from the vast existing body of knowledge available in percolation theory, thereby lending reliability and predictability.

The minimal plaquette model discussed Here evidently demonstrates the probability amplification achieved by GCP for smaller networks. For larger networks, like the other protocols presented in Table \ref{tab:per_thres}, we also employ numerics and finite-size scaling. The quantitative comparison clearly presents the fundamental theoretical advantage of GCP, when examined through the lens of the percolation threshold $\theta_T$.

\section{Conclusions}

Entanglement between two nodes is required for quantum communication in order to transfer, exchange, and convey quantum information. Quantum channels that exhibit entanglement will likely present an enhanced capacity for information transfer and improved security against interception. However, we require maximally entangled states to establish more secure and efficient quantum communication. As discussed earlier, creating and maintaining maximally entangled states is non-trivial. Due to various factors such as decoherence, photon loss, measurement errors, and other noise, only nonmaximally entangled states can be obtained. From a given network of nonmaximally entangled states, entanglement percolation enables the achievement of long-range maximally entangled states. It is natural to investigate the minimum strength of nonmaximally entangled states required for the possibility of entanglement percolation. Various protocols have been proposed for establishing long-range quantum communication based on percolation in statistical physics. However, developing a rather general protocol is of significant interest, as it emphasizes the fundamental principles underlying existing protocols rather than getting involved in the nuances of the individual existing protocols. Here, we focused exclusively on bipartite pure states. The present protocol, GCP, provides the lowest percolation threshold value. We also demonstrate how our protocol provides a lower percolation threshold by analytically solving a minimal model. However, in a quantum network, multipartite pure states can also arise. Additionally, mixed states are often encountered due to thermal noise and other random failures in quantum devices. Therefore, understanding the applicability and efficiency of our protocol in the context of multipartite pure entangled states and mixed states is also worth investigating. In conclusion, we can establish long-range quantum communication for a considerably lower amount of entanglement by utilizing the GCP protocol, compared to existing protocols.

\section{Acknowledgments}

We thank Somshubhro Bandyopadhyay and Ramnarayan Bera for their critical comments on the manuscript.

\appendix

\section{Calculation of the number and length of the shortest paths between two random nodes in a triangular lattice}
\label{sec:appendix_A}

For a given square lattice, as shown in Fig. \ref{fig:appendix}(a), let us consider node $O$ as the origin whose coordinates are $(0,0)$. The coordinates of the other nodes are shown in Fig. \ref{fig:appendix}(a). Let the coordinate of source $s$ and destination $t$ be $(x_s,y_s)$ and $(x_t,y_t)$. Without any loss of generality, for $x_s \le x_t$, the number of shortest paths between $s$ and $t$ is given by \cite{Malik2022}
\begin{equation}
n(s \to t)=\begin{pmatrix}
|x_s-x_t|+|y_s-y_t|\\
|x_s-x_t|
\end{pmatrix}
\label{eq:square_n}
\end{equation}

The length of the shortest path between $s=(x_s,y_s)$ and $t=(x_t,y_t)$ is, 
\begin{equation}
l(s \to t)=|x_s-x_t|+|y_s-y_t|
\label{eq:square_l}
\end{equation}

Obviously, $n(s \to t)=n(t \to s)$ and $l(s \to t)=l(t \to s)$. We can construct a triangular lattice from a given square lattice by simply adding an edge between $(i,j)$ and $(i+1,j+1)$, as shown in Fig. \ref{fig:appendix}(b). For instance, an edge will be added between $(0,0)$ and $(1,1)$. For $x_s \le x_t$, we can enumerate the following possibilities: 
\begin{figure}
      \centering
		\includegraphics[width=\linewidth, height=4cm]{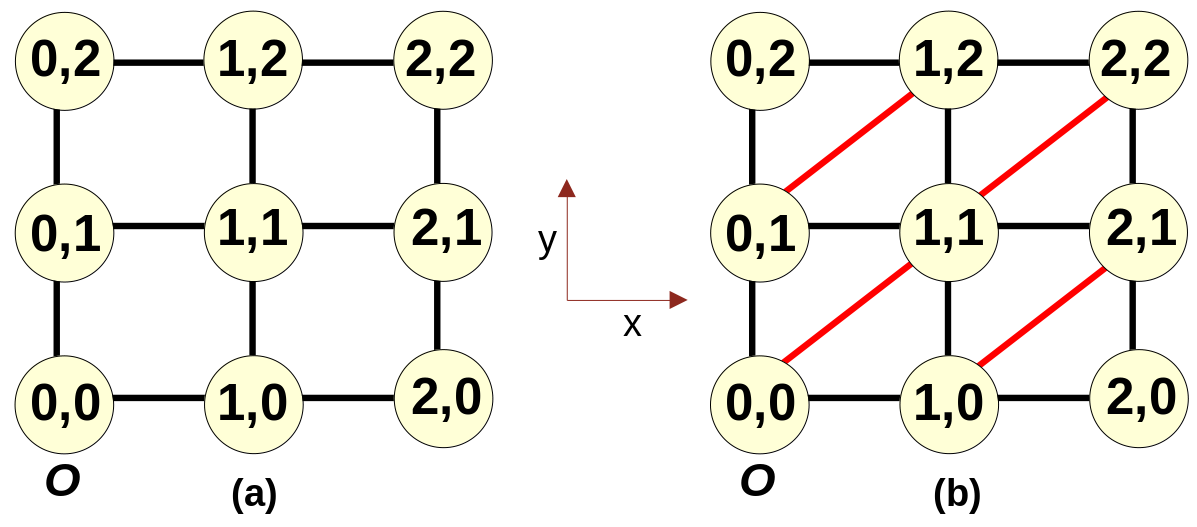}
		\caption{(a) Square lattice (L=3). (b) Triangular lattice (L=3). The edges in red are added to convert the square lattice into triangular.}
		\label{fig:appendix}
\end{figure}

\textbf{Case I:} $y_s \ge y_t$. $n(s \to t)$ and $l(s \to t)$ possess the same values as the square lattice, as shown in Eqs. \ref{eq:square_n} and \ref{eq:square_l}. 

\textbf{Case II:} $y_s < y_t$ and $s$ and $t$ are diagonal. From Fig. \ref{fig:appendix}(b), we observe $s=(0,0)$ and $t=(2,2)$ are diagonal. The only shortest path for diagonal nodes will be through the diagonal edges (highlighted in red). Therefore, the number of the shortest paths is $n=1$. For example, for $s=(0,0)$ and $t=(2,2)$, the only shortest path will be through $(1,1)$. There are no other possible shortest paths. However, the length of the shortest path will simply be the difference in either the $x$ coordinate or $y$ coordinate. For $s=(0,0)$ and $t=(2,2)$, the length of the shortest path $l=(2-0)=2$. If we denote $|x_s-x_t|$ and $|y_s-y_t|$ as $\Delta x$ and $\Delta y$, respectively, the length of the shortest path is
\begin{equation}
l(s \to t)=min\{\Delta x,\Delta y\}=\Delta x=\Delta y
\label{eq:triangular_l1}
\end{equation}

\textbf{Case III:} $y_s < y_t$. $s$ and $t$ are off-diagonal. In Fig. \ref{fig:appendix}(b), we observe  $s=(0,1)$ and $t=(2,2)$ are off-diagonal. Every shortest path between $s=(0,1)$ and $t=(2,2)$ will go through the diagonal edges [highlighted in red in Fig. \ref{fig:appendix}(b)]. Therefore, for each shortest path, there will be two parts: the diagonal and the off-diagonal parts. For the diagonal part, we have $l_1=min\{\Delta x,\Delta y\}$, similar to Eq. \ref{eq:triangular_l1}. For the off-diagonal part, we have $l_2=|\Delta x-\Delta y|$. The length of the shortest path is 
\begin{align}
l(s \to t) &=l_1 + l_2 \\
&=min\{\Delta x,\Delta y\}+|\Delta x-\Delta y|
\label{eq:triangular_l_off}
\end{align}

For example, let us calculate the number and length of the shortest paths between two non-neighboring nodes, $s$ and $t$, whose coordinates are $(x_s,y_s)=(0,1)$ and $(x_t,y_t)=(2,2)$, respectively. From Fig. \ref{fig:appendix}(a), we observe that we have two shortest paths between $s=(0,1)$ and $t=(2,2)$. These shortest paths are: $(0,1) \to (1,2) \to (2,2)$ and $(0,1) \to (1,1) \to (2,2)$. Both of these shortest paths consist of two parts: the diagonal (highlighted by red solid lines) and the off-diagonal (highlighted by black solid lines) parts. In the case of shortest path $(0,1) \to (1,2) \to (2,2)$, the diagonal part is $(0,1) \to (1,2)$ and the off-diagonal part is $(1,2) \to (2,2)$, as seen in Fig. \ref{fig:appendix}(b). Similary, for another shortest path, $(0,1) \to (1,1) \to (2,2)$, the diagonal and off-diagonal parts are $(1,1) \to (2,2)$ and $(0,1) \to (1,2)$, respectively, shown in Fig. \ref{fig:appendix}(b). We can calculate the length of these diagonal and off-diagonal parts separately. The length of the diagonal part can be evaluated as, $l_1=min\{\Delta x,\Delta y\}=min\{|x_s-x_t|,|y_s-y_t|\}=min\{|0-2|,|1-2|\}=1$. On the other hand, the length of the off-diagonal part can be estimated as $l_2=\Delta x-\Delta y=|x_s-x_t|-|y_s-y_t|=|0-2|-|1-2|=2-1=1$. The total length of the shortest path between $s=(0,1)$ and $t=(2,2)$ will be, $l=l_1+l_2=1+1=2$. On the other hand, the number of shortest paths is $n=\Mycomb[l]{l_1}=\Mycomb[2]{1}=2$. In general, 
\begin{align}
n(s \to t)=\Mycomb[l]{l_1}=\begin{pmatrix}
min\{\Delta x,\Delta y\}+|\Delta x-\Delta y|\\
min\{\Delta x,\Delta y\}
\end{pmatrix}
\label{eq:triangular_n_off}
\end{align}

\section{Implementation of the GCP Protocol}
\label{sec:appendix_B}

The Monte Carlo simulation of the GCP protocol for a given lattice of size $N$ can be executed per the following outline.

\begin{enumerate}
\item A 2D lattice (square, triangular, or hexagonal) with $N$ nodes is generated with specified boundary conditions.

\item For every unique pair of non-neighboring nodes $(i, j)$ in the network, the number $n_{ij}$ and length $l_{ij}$ of all shortest paths are determined using Dijkstra's algorithm (or via exact combinatorics, as detailed for specific lattices in  Appendix \ref{sec:appendix_A}).

\item Using the series rule [Eq. \ref{eq:series_rule_c}] and parallel rule [Eq. \ref{eq:parallel_rule_c}], the effective concurrence is calculated for every node pair. It is used to compute the singlet conversion probability $p_{{}_{\text{GCP}}}^{ij}$ between nodes $i$ and $j$.

\item For neighboring nodes, we calculate the singlet conversion probability using CEP, i.e., $p_{{}_{\text{CEP}}}^{ij}$.

\item For a specific value of initial entanglement $\theta$, a uniform random number $r \in [0, 1]$ is generated for each pair $(i, j)$. If $r \le p^{ij}$, a singlet (an undirected graph edge) is successfully established between $i$ and $j$.

\item The well-known Newman-Ziff algorithm \cite{Newman2001} for labeling connected components is applied to the resulting unweighted graph of singlets to identify the size of the largest connected component, $s_{max}$.

\item Steps 4 and 5 are repeated across $E_N = 1000$ independent statistical ensembles to compute the average density of the spanning cluster, $P \equiv f_{gc}$.

\end{enumerate}

This process can be repeated across a range of different values of $\theta$  at various $N$ followed by data collapse and the extraction of the critical threshold $\theta_T$.

{\tiny }

\bibliographystyle{biblatex}

\bibliographystyle{apsrev4-2}

\bibliography{GCP}

@inproceedings{Buhrman1998,
  title={Quantum vs. classical communication and computation},
  author={Buhrman, Harry and Cleve, Richard and Wigderson, Avi},
  booktitle={Proceedings of the 30th annual ACM symposium on Theory of computing, New York},
  pages={63--68},
  year={1998}
}

@article{Flamini2018,
  title={Photonic quantum information processing: a review},
  author={Flamini, Fulvio and Spagnolo, Nicolo and Sciarrino, Fabio},
  journal={Rep. Prog. Phys.},
  volume={82},
  number={1},
  pages={016001},
  year={2018},
  publisher={IOP Publishing}
}

@article{Gisin2007,
  title={Quantum communication},
  author={Gisin, Nicolas and Thew, Rob},
  journal={Nat. Photonics},
  volume={1},
  number={3},
  pages={165--171},
  year={2007},
  publisher={Nature Publishing Group UK London}
}

@article{Horodecki2009,
  title = {Quantum entanglement},
  author = {Horodecki, Ryszard and Horodecki, Pawe\l{} and Horodecki, Micha\l{} and Horodecki, Karol},
  journal = {Rev. Mod. Phys.},
  volume = {81},
  issue = {2},
  pages = {865--942},
  numpages = {0},
  year = {2009},
  month = {Jun},
  publisher = {American Physical Society},
}

@article{Wootters1998,
  title={Quantum entanglement as a quantifiable resource},
  author={Wootters, William K},
  journal={Philos. Trans. Math. Phys. Eng. Sci.},
  volume={356},
  number={1743},
  pages={1717--1731},
  year={1998},
  publisher={The Royal Society}
}

@article{Penrose1998,
  title={Quantum computation, entanglement and state reduction},
  author={Penrose, Roger},
  journal={Philos. Trans. Math. Phys. Eng. Sci.},
  volume={356},
  number={1743},
  pages={1927--1939},
  year={1998},
  publisher={The Royal Society}
}

@article{Bennett1993,
  title = {Teleporting an unknown quantum state via dual classical and Einstein-Podolsky-Rosen channels},
  author = {Bennett, Charles H. and Brassard, Gilles and Cr\'epeau, Claude and Jozsa, Richard and Peres, Asher and Wootters, William K.},
  journal = {Phys. Rev. Lett.},
  volume = {70},
  issue = {13},
  pages = {1895--1899},
  numpages = {0},
  year = {1993},
  month = {Mar},
  publisher = {American Physical Society}
}

@article{Hu2025,
  author  = {Hu, Xinqi and Dong, Gaogao and Christensen, Kim and Sun, Hanlin and Fan, Jingfang and Tian, Zihao and Gao, Jianxi and Havlin, Shlomo and Lambiotte, Renaud and Meng, Xiangyi},
  title   = {Unveiling the importance of nonshortest paths in quantum networks},
  journal = {Sci. Adv.},
  volume  = {11},
  number  = {9},
  pages   = {eadt2404},
  year    = {2025}
}

@article{Yin2020,
  title={Entanglement-based secure quantum cryptography over 1,120 kilometres},
  author={Yin, Juan and Li, Yu Huai and Liao, Sheng Kai and Yang, Meng and Cao, Yuan and Zhang, Liang and Ren, Ji Gang and Cai, Wen Qi and Liu, Wei Yue and Li, Shuang Lin and others},
  journal={Nature},
  volume={582},
  number={7813},
  pages={501--505},
  year={2020},
  publisher={Nature Publishing Group UK London}
}

@article{Watts2021,
  title={Photon quantum entanglement in the MeV regime and its application in PET imaging},
  author={Watts, DP and Bordes, J and Brown, JR and Cherlin, A and Newton, R and Allison, J and Bashkanov, M and Efthimiou, N and Zachariou, NA},
  journal={Nat. Commun.},
  volume={12},
  number={1},
  pages={2646},
  year={2021},
  publisher={Nature Publishing Group UK London}
}

@article{Degen2017,
  title = {Quantum sensing},
  author = {Degen, C. L. and Reinhard, F. and Cappellaro, P.},
  journal = {Rev. Mod. Phys.},
  volume = {89},
  issue = {3},
  pages = {035002},
  numpages = {39},
  year = {2017},
  month = {Jul},
  publisher = {American Physical Society}
}

@article{Ekert1998,
  title={Quantum algorithms: entanglement--enhanced information processing},
  author={Ekert, Artur and Jozsa, Richard},
  journal={Philos. Trans. Math. Phys. Eng. Sci.},
  volume={356},
  number={1743},
  pages={1769--1782},
  year={1998},
  publisher={The Royal Society}
}

@article{Perseguers2013,
  title={Distribution of entanglement in large-scale quantum networks},
  author={Perseguers, S{\'e}bastien and Lapeyre, GJ and Cavalcanti, D and Lewenstein, M and Ac{\'\i}n, A},
  journal={Rep. Prog. Phys.},
  volume={76},
  number={9},
  pages={096001},
  year={2013},
  publisher={IOP Publishing}
}

@article{Simon2017,
  title={Towards a global quantum network},
  author={Simon, Christoph},
  journal={Nat. Photonics},
  volume={11},
  number={11},
  pages={678--680},
  year={2017},
  publisher={Nature Publishing Group UK London}
}

@article{Brito2020,
  title={Statistical properties of the quantum internet},
  author={Brito, Samura{\'\i} and Canabarro, Askery and Chaves, Rafael and Cavalcanti, Daniel},
  journal={Phys. Rev. Lett.},
  volume={124},
  number={21},
  pages={210501},
  year={2020},
  publisher={APS}
}

@article{Schlosshauer2019,
  title={Quantum decoherence},
  author={Schlosshauer, Maximilian},
  journal={Phys. Rep.},
  volume={831},
  pages={1--57},
  year={2019},
  publisher={Elsevier}
}

@article{Ekert1993,
  title = "{``Event-ready-detectors'' Bell experiment via entanglement swapping}",
  author = {\ifmmode \dot{Z}\else \.{Z}\fi{}ukowski, M. and Zeilinger, A. and Horne, M. A. and Ekert, A. K.},
  journal = {Phys. Rev. Lett.},
  volume = {71},
  issue = {26},
  pages = {4287--4290},
  numpages = {0},
  year = {1993},
  month = {Dec},
  publisher = {American Physical Society}
}

@article{Bose1999,
  title = {Purification via entanglement swapping and conserved entanglement},
  author = {Bose, S. and Vedral, V. and Knight, P. L.},
  journal = {Phys. Rev. A},
  volume = {60},
  issue = {1},
  pages = {194--197},
  numpages = {0},
  year = {1999},
  month = {Jul},
  publisher = {American Physical Society}
}

@article{Acin2007,
  title={Entanglement percolation in quantum networks},
  author={Ac{\'\i}n, Antonio and Cirac, J Ignacio and Lewenstein, Maciej},
  journal={Nat. Phys.},
  volume={3},
  number={4},
  pages={256--259},
  year={2007},
  publisher={Nature Publishing Group UK London}
}

@article{Shchukin2022,
  title={Optimal entanglement swapping in quantum repeaters},
  author={Shchukin, Evgeny and van Loock, Peter},
  journal={Phys. Rev. Lett.},
  volume={128},
  number={15},
  pages={150502},
  year={2022},
  publisher={APS}
}

@article{Meng2021,
  title={Concurrence percolation in quantum networks},
  author={Meng, Xiangyi and Gao, Jianxi and Havlin, Shlomo},
  journal={Phys. Rev. Lett.},
  volume={126},
  number={17},
  pages={170501},
  year={2021},
  publisher={APS}
}

@article{Malik2022,
  title={Concurrence percolation threshold of large-scale quantum networks},
  author={Malik, Omar and Meng, Xiangyi and Havlin, Shlomo and Korniss, Gyorgy and Szymanski, Boleslaw Karol and Gao, Jianxi},
  journal={Commun. Phys.},
  volume={5},
  number={1},
  pages={193},
  year={2022},
  publisher={Nature Publishing Group UK London}
}

@article{Perseguers2008,
  title={Entanglement distribution in pure-state quantum networks},
  author={Perseguers, S{\'e}bastien and Cirac, J Ignacio and Ac{\'\i}n, Antonio and Lewenstein, Maciej and Wehr, Jan},
  journal={Phys. Rev. A},
  volume={77},
  number={2},
  pages={022308},
  year={2008},
  publisher={APS}
}

@article{Cuquet2009,
  title={Entanglement percolation in quantum complex networks},
  author={Cuquet, Mart{\'\i} and Calsamiglia, John},
  journal={Phys. Rev. Lett.},
  volume={103},
  number={24},
  pages={240503},
  year={2009},
  publisher={APS}
}

@article{Perseguers2010,
  title={Multipartite entanglement percolation},
  author={Perseguers, S{\'e}bastien and Cavalcanti, D and Lapeyre Jr, GJ and Lewenstein, M and Ac{\'\i}n, A},
  journal={Phys. Rev. A},
  volume={81},
  number={3},
  pages={032327},
  year={2010},
  publisher={APS}
}

@book{Stauffer1994,
  title = {Introduction to percolation theory},
  author = {Stauffer, D. and Aharony, A.},
  year = {1994},
  publisher = {CRC Press},
  address = {Boca Raton, FL}
}

@article{Dijkstra1959,
  author  = {Dijkstra, E. W.},
  title   = {A note on two problems in connexion with graphs},
  journal = {Numerische Mathematik},
  year    = {1959},
  volume  = {1},
  number  = {1},
  pages   = {269--271},
  publisher = {Springer}
}

@book{Binder1992,
  title={Monte Carlo Simulation in Statistical Physics},
  author={Binder, Kurt},
  year={1992},
  publisher={Springer-Verlag, New York}
}

@article{Vicente2024,
  title = {Maximally Entangled Mixed States for a Fixed Spectrum Do Not Always Exist},
  author = {de Vicente, Julio I.},
  journal = {Phys. Rev. Lett.},
  volume = {133},
  issue = {5},
  pages = {050202},
  numpages = {5},
  year = {2024},
  month = {Jul},
  publisher = {American Physical Society}
}

@article{Neumann2018,
  title="{Q3Sat: quantum communications uplink to a 3U CubeSat—feasibility \& design}",
  author={Neumann, Sebastian Philipp and Joshi, Siddarth Koduru and Fink, Matthias and Scheidl, Thomas and Blach, Roland and Scharlemann, Carsten and Abouagaga, Sameh and Bambery, Daanish and Kerstel, Erik and Barthelemy, Mathieu and others},
  journal={EPJ Quantum Technology},
  volume={5},
  number={1},
  pages={1--24},
  year={2018},
  publisher={SpringerOpen}
}

@article{Zuo2021,
  title={Overcoming the uplink limit of satellite-based quantum communication with deterministic quantum teleportation},
  author={Zuo, Zhiyue and Wang, Yijun and Liao, Qin and Guo, Ying},
  journal={Phys. Rev. A},
  volume={104},
  number={2},
  pages={022615},
  year={2021},
  publisher={APS}
}

@article{Brito2021,
  title={Satellite-based photonic quantum networks are small-world},
  author={Brito, Samura{\'\i} and Canabarro, Askery and Cavalcanti, Daniel and Chaves, Rafael},
  journal={PRX Quantum},
  volume={2},
  number={1},
  pages={010304},
  year={2021},
  publisher={APS}
}

@book{malthe2024percolation,
  author    = {Anders Malthe-S{\o}renssen},
  title     = {Percolation Theory Using Python},
  publisher = {Springer},
  address   = {Cham},
  year      = {2024},
}

@article{Gimenez2025,
  author  = {M. C. Gimenez and L. Reinaudi and P. M. Centres},
  title   = {Percolation threshold and critical exponent analysis in equilibrium systems on simple cubic and BCC lattices},
  journal = {Physica A: Statistical Mechanics and its Applications},
  volume  = {668},
  pages   = {130562},
  year    = {2025}
}

@article{Newman2001,
  author  = {M. E. J. Newman and R. M. Ziff},
  title   = {Fast Monte Carlo algorithm for site or bond percolation},
  journal = {Phys. Rev. E},
  volume  = {64},
  pages   = {016706},
  year    = {2001}
}

@misc{gcp_code,
  note = {Code for: General Concurrence Percolation on Quantum Networks, https://github.com/deepnath441/GCP}
}

\end{document}